\documentclass[apj]{emulateapj}
\usepackage{apjfonts}
\usepackage{natbib}

\DeclareGraphicsExtensions{.eps}

\newcommand{\alf}{$\alpha_{\rm 3000}$}
\newcommand{\alfe}{\alpha_{\rm 3000}}

\newcommand{\figu}{Figure~}

\newcommand{\sect}{Section~}

\newcommand{\ergse}{{erg~s^{-1}}}

\newcommand{\mbh}{$M_{\rm bh}$}
\newcommand{\mbhe}{M_{\rm bh}}

\newcommand{\msune}{M_{\odot}}

\newcommand{\ang}{{\AA}}

\begin{document}

\voffset=-0.50in

\def\sarc{$^{\prime\prime}\!\!.$}
\def\arcsec{$^{\prime\prime}$}
\def\arcmin{$^{\prime}$}
\def\degr{$^{\circ}$}
\def\seco{$^{\rm s}\!\!.$}
\def\ls{\lower 2pt \hbox{$\;\scriptscriptstyle \buildrel<\over\sim\;$}}
\def\gs{\lower 2pt \hbox{$\;\scriptscriptstyle \buildrel>\over\sim\;$}}

\title{The optical-UV emissivity of quasars: dependence on black hole mass and radio loudness}

\author{Francesco Shankar\altaffilmark{1}, Giorgio Calderone\altaffilmark{2}, Christian Knigge\altaffilmark{1}, James Matthews\altaffilmark{1}, Rachel Buckland\altaffilmark{1}, Krzysztof Hryniewicz\altaffilmark{3}, Gregory Sivakoff\altaffilmark{4}, Xinyu Dai\altaffilmark{5}, Kayleigh Richardson \altaffilmark{1}, Jack Riley\altaffilmark{1}, James Gray\altaffilmark{1},
Fabio La Franca\altaffilmark{6}, Diego Altamirano\altaffilmark{1}, Judith Croston\altaffilmark{1}, Poshak Gandhi\altaffilmark{1}, Sebastian H\"{o}nig\altaffilmark{1}, Ian McHardy\altaffilmark{1}, Matthew Middleton\altaffilmark{7}}
\altaffiltext{1}{Department of Physics and Astronomy, University of Southampton, Highfield, SO17 1BJ, UK; F.Shankar@soton.ac.uk}
\altaffiltext{2}{INAF--Osservatorio Astronomico di Brera, via E. Bianchi 46, I-23807 Merate (LC), Italy; calderone@oats.inaf.it}
\altaffiltext{3}{ISDC Data Centre for Astrophysics, Observatoire de Gen\`{e}ve, Universit\'{e} de Gen\`{e}ve, Chemin d'Ecogia 16, 1290 Versoix, Switzerland}
\altaffiltext{4}{Department of Physics, University of Alberta, CCIS 4-183, Edmonton, Alberta T6G 2E1, Canada}
\altaffiltext{5}{Homer L. Dodge Department of Physics and Astronomy, University of Oklahoma, Norman, OK 73019, USA}
\altaffiltext{6}{Dipartimento di Matematica e Fisica, Universit\'{a} degli Studi Roma Tre, via della Vasca Navale 84, 00146 Roma, Italy}
\altaffiltext{7}{Institute of Astronomy, Madingley Road, Cambridge CB3 0HA, UK}
\begin{abstract}
We analyzed a large sample of radio-loud and radio-quiet quasar
spectra at redshift $1.0 \le z \le 1.2$ to compare the inferred
underlying quasar continuum slopes (after removal of the host galaxy
contribution) with accretion disk models. The latter
predict redder (decreasing) \alf\ continuum slopes ($L_{\nu}
\propto \nu^{\alpha}$ at $3000$\ang) with increasing black hole mass,
bluer \alf\ with increasing luminosity at 3000\ang, and bluer \alf\ with increasing spin of the black hole, when all
other parameters are held fixed.  We find no clear evidence for any of these predictions in the data. In particular we find that: (i) \alf\ shows no significant dependence on black hole mass or luminosity. Dedicated Monte Carlo tests suggest that the substantial observational uncertainties in the black hole virial masses can effectively erase any intrinsic dependence of \alf\ on black hole mass, in line with some previous studies. (ii) The mean slope \alf\ of radio-loud sources, thought to be produced by rapidly spinning black holes, is comparable to, or even \emph{redder} than, that of radio-quiet quasars. Indeed, although quasars appear to become more radio loud with decreasing luminosity, we still do not detect any significant dependence of \alf\ on radio loudness.
The predicted mean \alf\ slopes tend to be bluer than in the data. Disk models with high inclinations and dust extinction tend to produce redder slopes closer to empirical estimates.
Our mean \alf\ values are close to the ones independently inferred at $z<0.5$ suggesting weak evolution with redshift, at least for moderately luminous quasars.
\end{abstract}


\section{Introduction}
\label{sec|intro}

The current accretion paradigm on compact objects (from stellar
sources to super-massive black holes) assumes that infalling matter
will settle in a rotating disk in which the gas looses its angular
momentum via dynamical friction.  The disk is usually assumed to be
geometrically thin and optically thick \citep{SSdisk}, i.e., each
annulus radiates as a black body whose temperature depends on its
radius.  The resulting spectrum shows a broad peak at a characteristic
energy related to the temperature at the inner radii of the disk.
In Active Galactic Nuclei (AGN) the disk spectrum peaks in the
UV/soft X-rays \citep[the ``Big Blue Bump'', e.g.,][]{FKR}.

The disk model spectra are self--similar in
logarithmic $\nu L_\nu$ plots, and the location of the peak depends on
the black hole mass, the accretion rate, and the location of the innermost
stable circular orbit ($R_{\rm ISCO}$, related to black hole spin).  In particular, at fixed accretion rate and $R_{\rm
  ISCO}$, the peak will shift to lower frequencies with increasing
black hole mass.  Since the peak is expected to lie at wavelengths
$\lesssim$~3000\ang, the peak shift results in a reddening (decrease) of the
spectral slope at 3000\ang\ (\alf, Eq. \ref{eq|alpha}).  Also, at fixed
black hole mass and spin, the peak will shift to higher frequencies
with increasing UV luminosity, and the spectral slope \alf\ will
become bluer.  Finally, at fixed black hole mass and accretion rate, the
peak will shift to higher frequencies with decreasing $R_{\rm ISCO}$
(i.e. increasing black hole spin), and the slope at UV wavelengths is
again expected to become bluer \citep[see, e.g., \figu2 in][]{DavisLaor11}.

According to some popular models
\citep[e.g.,][]{BlandfordZ}, powerful jets are
produced by rapidly spinning black holes.  In this framework, the spectral
properties of radio-loud and radio-quiet quasars are expected to show significant
differences, and may provide valuable insights for testing
theoretical expectations.

\begin{figure*}
    \center{\includegraphics[width=18truecm]{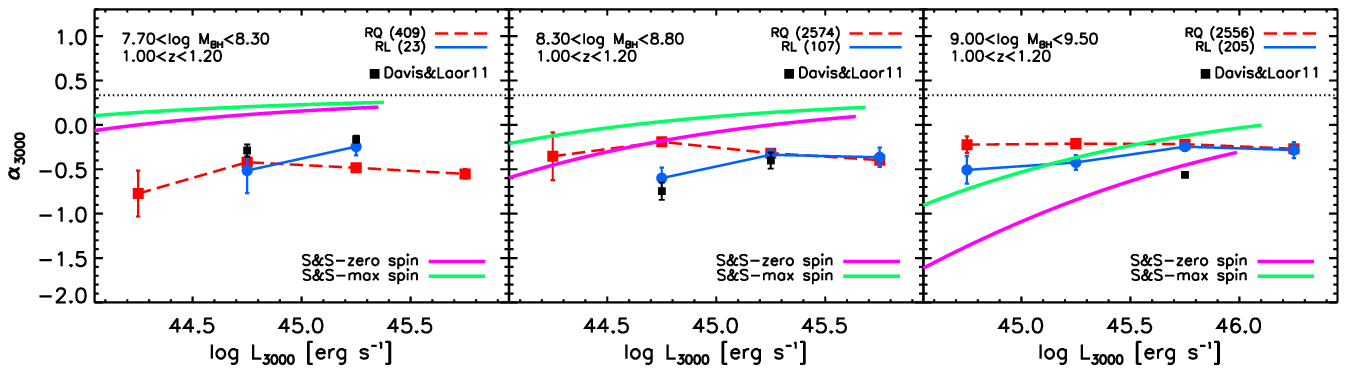}
    \caption{Mean \alf\ as a function of the luminosity at 3000\ang\ (rest frame) for black holes in the range $7.7<\log \mbhe/\msune<8.3$ (left), $8.3<\log \mbhe/\msune<8.8$ (middle), and $9.0<\log \mbhe/\msune<9.5$ (right). The solid, purple lines mark the prediction of the \citet{SSdisk} disk model for a non-rotating, zero spin black hole with mass $\log \mbhe/\msune=8.0, 8.5, 9.25$ for the left, middle, and right panels, respectively. The solid, green lines show the results of the same blackbody models for maximally spinning black holes. Models are plotted up to the Eddington limit corresponding to the mean black hole mass labeled in each panel. All radio-loud sources
      tend to have \alf\ comparable to those of radio-quiet quasars at the highest luminosity bins, i.e., $\log L_{\rm 3000}/{\rm \ergse} > 45$,
      and systematically redder at progressively lower luminosities.
     The horizontal dotted lines mark the canonical accretion disk slope at wavelengths significant
longer than the peak wavelength for a standard accretion disk model. The filled black squares are data derived by \citet[][]{DavisLaor11} from a sample of optical quasars at $z<0.5$.
    \label{fig|AlphaLoptMbh}}}
\end{figure*}

The main purpose of this letter is to compare the AGN UV spectral slopes
with predictions of standard thin-disk models using an improved method
to measure the quasar continuum applied to a very large sample.

\section{Data}
\label{sec|method}

The quasar sample used in this work is a subset of the SDSS DR7 quasar
sample discussed in \citet{Schneider10}.  The spectral properties of
this sample are discussed in the catalog by \citet{Shen11}.  However,
we re-analyzed all sources with $z<2$ to better
account for the host galaxy and the iron complex contributions on each
spectrum, and to identify the actual broad band AGN continuum.  The
details will be presented in a forthcoming paper (Calderone et
al. 2016, in preparation), but we provide a brief description here: the
SDSS DR10 spectrum of each source is fitted via a model that includes
the quasar continuum (modeled as a smoothly broken power law), an
elliptical host galaxy template \citep{Mannucci01},
an optical \citep{Veron04} and UV
\citep{Veste01} iron template, and a list of common
broad and narrow emission lines.  The smoothly broken power law
extends over the entire observed wavelength range, providing a
reasonable estimate of the actual AGN continuum. The host galaxy template is chosen to
be that of an elliptical galaxy since such galaxies are those usually observed to
be the hosts of massive, luminous black holes, based on direct photometric studies
\citep[e.g.,][]{Falomo14} and local scaling relations
\citep[e.g.,][]{ShankarReview,KormendyHo}.

Our catalog thus contains updated continuum (isotropic) luminosities and slopes
with respect to the \citet{Shen11} one, where the host galaxy and iron
contributions were neglected. Note that our luminosity and slopes estimates
are measured on the smoothly broken power law component that extends
over the whole rest-frame observed wavelength range. This implies that the slopes
are evaluated on a wider wavelength range with respect to \citet{Shen11}.

The purpose of our spectral analysis is to estimate the underlying quasar
continuum slope \alf\ defined as follows:
\begin{equation}
\alfe=\frac{d\log L_{\nu}}{d\log \nu}\, ,
    \label{eq|alpha}
\end{equation}
calculated at $3000$\ang\ (source rest frame).

In the following we will consider several sub--samples of the original
\citet{Schneider10} sample according to either: the virial black hole
mass as listed in \citet{Shen11} derived mainly from MgII in the redshift range of interest to this work;
the continuum luminosity and slopes at 3000\ang\ (rest frame) according to the results of our new spectral
analysis; and the radio loudness $R=f_{\rm 6 cm}/f_{\rm 2500}$
\citep{Jiang07} defined as the ratio between the flux density
($f_{\rm \nu}$) at 6 cm and 2500\ang\ rest-frame respectively, available for all SDSS quasars matched against the Faint
Images of the Radio Sky at Twenty-Centimeters (FIRST) survey at 1.4
GHz \citep{FIRST}. We devote specific attention to the difference in spectral slopes
between radio-quiet quasars, lying within the footprint of FIRST but with $R=0$,
and very radio-loud sources with $R>100$, though the exact threshold chosen for radio loudness
does not alter any of our conclusions. We also briefly mention results based on ``radio morphology'' as reported in
\citet{Shen11}, in which radio-loud quasars are broadly divided into ``FR1'' and
``FR2'' sources \citep{FR74}, i.e. core-- or lobe--dominated (see \citealt{Jiang07} for details).

We will specifically select sources lying in the redshift interval $1.0<z<1.2$.
This redshift window allows for more accurate measurements of the SDSS
optical-UV quasar continuum spectra around 3000\ang. As detailed in Calderone et al. (2016) in fact,
spectral slopes measured with our procedure on quasars with $z<1.0$ or $z>1.3$
may be biased due to the limited SDSS wavelength spectral coverage.
Finally, Malmquist bias effects and reddening/extinction by intervening cosmic
dust \citep[][]{XieSED} could become increasing issues at higher redshifts and/or larger redshift bins, further biasing the true distributions of \alf.

\section{Disk models}
\label{sec|diskmodel}

\begin{figure*}
    \center{\includegraphics[width=15truecm]{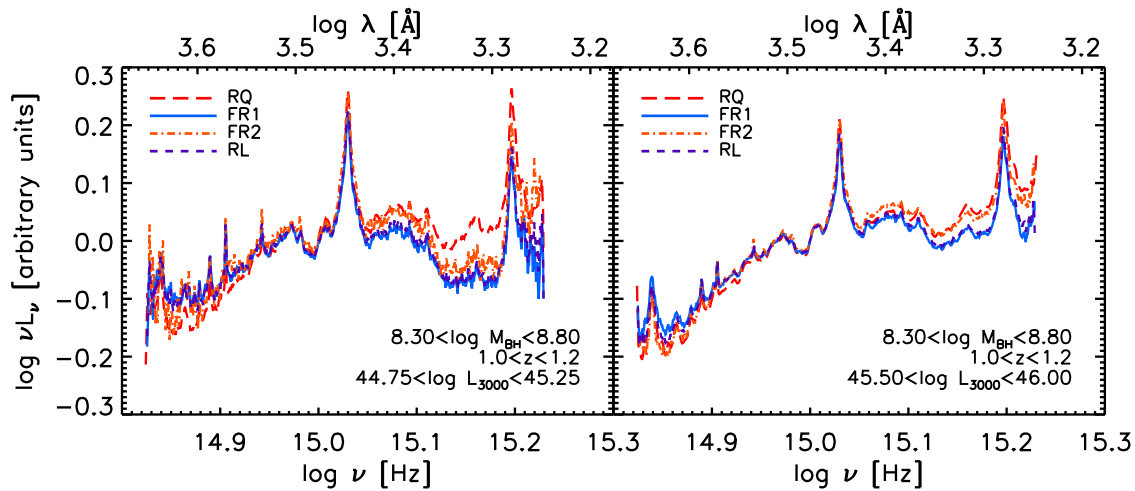}
    \caption{Stacked SDSS spectra of quasars with black hole masses in the range $8.3<\log \mbhe/\msune<8.8$, in the redshift window $1.0<z<1.2$, and normalized at $3000$\ang. We include radio-quiet, core-dominated FR1, and lobe-dominated FR2 radio sources, as well as radio-loud quasars with $R>10$, for two intervals of $L_{\rm 3000}$ luminosity, as labeled. It is apparent that all radio-loud sources, irrespective of their exact classifications, show flatter profiles than radio-quiet quasars.
    \label{fig|Stacked}}}
\end{figure*}

We compare our data with
predictions of the non-relativistic, steady-state, geometrically
thin, optically thick standard accretion disk \citet{SSdisk}.
In brief, the model first computes the amount of gravitational energy $\zeta(R)$
released from each disk annulus of size $2\pi RdR$ and, under
the assumption of optical thickness, it converts it into a blackbody
temperature $T(R) \propto \zeta(R)^{1/4}$. The expected
frequency-dependent luminosity $L_{\nu}$ is then computed as a
superposition of blackbody spectra $B[\nu,T(R)]$.
The physical input parameters in the standard disk model are the
innermost stable circular orbit $R_{\rm isco}$, directly linked to the spin/radiative efficiency,
the black hole mass \mbh, and the mass accretion rate $\dot{M}_{\rm
  acc}$ \citep[e.g.,][and references therein]{Thorne74,Calderone13}.

We also compare our main observational results with AGNSPEC \citep{Hubeny97,Hubeny00,Hubeny01}, a self-consistent numerical model for AGN disk spectra as viewed by an observer at a given angle with respect to the axis of rotation. The model properly considers departures from local thermodynamic equilibrium, and takes into account the vertical structure of the disk, as well as relativistic Doppler shifts, gravitational redshifts and light bending in a rotating spacetime \citep{Hubeny00}.
The direct comparison between the predictions from basic disk models and AGNSPEC will allow us to pin down which additional features from the latter (more realistic) models are essential in describing the UV spectra of quasars.

For completeness, all model luminosities are converted to ``isotropic equivalent'' \citep[e.g.,][]{Calderone13} via $1+\cos i$ with $i$ the maximum viewing angle. Unless otherwise noted, in the following we will assume $i=45^{\circ}$ \citep[e.g.,][]{UrryPadovani}.

\section{Results}
\label{sec|results}

\figu\ref{fig|AlphaLoptMbh} shows the mean (and standard deviation of the mean) \alf\ as a function of the
luminosity at 3000\ang\ (rest frame) for black holes in the range $7.7<\log \mbhe/\msune<8.3$ (left), $8.3<\log \mbhe/\msune<8.8$ (middle), and $9.0<\log \mbhe/\msune<9.5$ (right). We chose bins (in black hole mass and/or luminosity) sufficiently large to enhance statistics but still small enough to provide multiple bins across our parameter space. We note that none of our results depend on the exact binning. In this and all subsequent figures, the total numbers of radio-quiet (red, long-dashed lines) and radio-loud (blue, solid lines) sources is indicated in the legend of each plot. All our data gather around rather constant mean slopes of $-0.5<\alfe<-0.3$, quite independently of black hole mass, luminosity, or the quasar radio state. Most of the radio-loud sources share \alf\ slopes very close to those of radio-quiet sources of similar black hole mass and luminosity. At luminosities below $\log L_{\rm 3000}/{\rm \ergse} \lesssim 45$ radio sources may tend towards even redder spectra (lower \alf). We also note that the mean \alf\ values extracted from our data agree well with those derived at $z<0.5$ by \citet[][]{DavisLaor11} for a sample of 80 PG measured in the interval between 4816\ang\ and 1549\ang, at least at lower luminosities, suggesting little redshift evolution in the optical-UV spectral properties of quasars.

We compare our measurements with the predictions of the \citet{SSdisk} disk model for a non-rotating, zero spin (red lines) and maximally spinning (green lines) black holes of same mass and luminosity. As already emphasized by a number of previous groups \citep[e.g.,][and references therein]{Bonning07,Davis07,LD14}, the results of direct quasar spectral fittings are often at variance with the theoretical expectations of the thin disk model. Our measurements further reinforce this by pointing to rather constant mean \alf\ values, in apparent conflict with basic predictions from standard accretion disk theory that requires a significant drop of \alf\ at fixed optical luminosity when moving from low to high-mass black holes, as evident from the models reported in \figu\ref{fig|AlphaLoptMbh} when moving from the left to the right panel.

The differences in \alf\ between radio-loud and radio-quiet quasars, in particular, are not an artifact of the exact spectral fitting procedure and/or on the exact definition of radio loudness. Indeed, such systematic differences are already clearly evident directly on the stacked spectra. \figu\ref{fig|Stacked} reports
the stacked spectra of radio-quiet and radio-loud quasars with black hole masses in the range $8.3<\log \mbhe/\msune<8.8$ at $1.0<z<1.2$, and normalized at $3000$\ang.
Irrespective of the exact subsample of radio sources, either FR1, FR2, or simply based on their radio loudness (here we report sources with $R>10$ to increase the statistics), all radio-detected sources show systematically redder profiles (lower \alf) with respect to their radio-quiet counterparts. This difference is however more marked for less luminous sources (left panel), while it progressively disappears for brighter sources (right panel), in agreement with \figu\ref{fig|AlphaLoptMbh}.

\begin{figure*}
    \center{\includegraphics[width=15truecm]{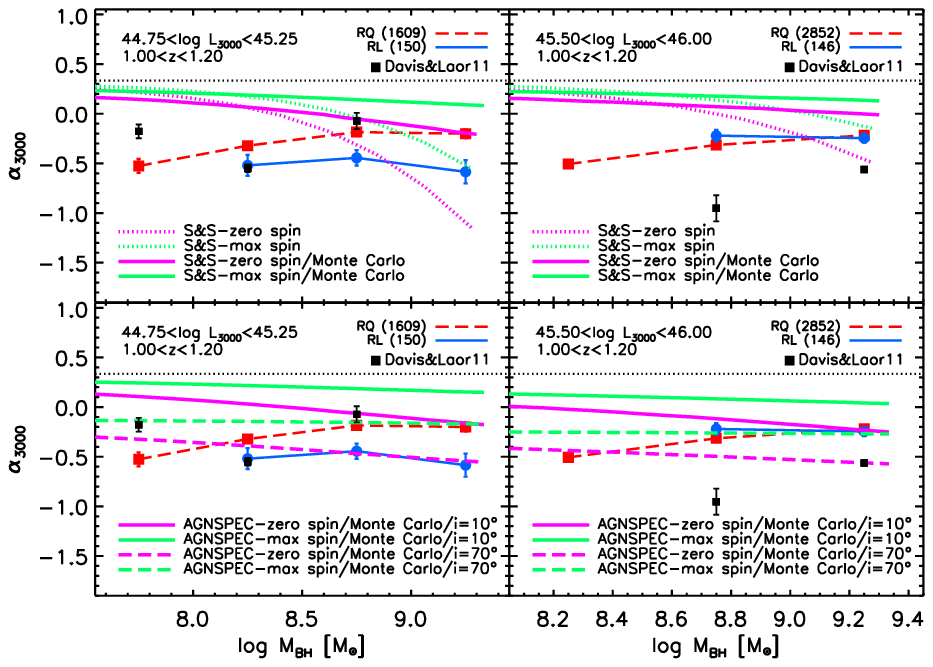}
    \caption{Mean \alf\ for radio-quiet and radio-loud sources as a function of black hole mass in two different luminosity bins, $44.75<\log L_{\rm 3000}/\ergse<45.25$ (left) and $45.5<\log L_{\rm 3000}/\ergse<46.0$ (right), as labeled. The data in the upper and lower panels are compared with \citet{SSdisk} and AGNSPEC disk models, respectively (see \sect\ref{sec|diskmodel}). The magenta and green lines in all panels refer to models with zero and maximum spin respectively.
    The dotted and solid lines in the upper panels refer, respectively, to the raw model outputs and to the Monte Carlo simulations inclusive of statistical errors in black hole mass and luminosity (see text for details). The solid and long-dashed lines in the lower panels instead are from Monte Carlo simulations with low ($10^{\circ}$) and extreme inclinations ($70^{\circ}$), respectively.
    \label{fig|AlphaMonteCarlo}}}
\end{figure*}

\figu\ref{fig|AlphaMonteCarlo} further compares model predictions against SDSS data by specifically plotting the behavior of \alf\ as a function of black hole mass for quasars in the narrow bins of optical luminosity $44.75<\log L_{\rm 3000}/{\rm \ergse}<45.25$ (left) and $45.5<\log L_{\rm 3000}/{\rm \ergse}<46.0$ (right). Our measured \alf\ is rather flat, or even slightly bluer with increasing black hole mass in the range $-0.6<\alfe<-0.2$, in nice agreement with the values for $\alpha$ between $2200-4000\, {\AA}$ inferred by \citet[][their \figu3]{Davis07}, who also emphasized very weak dependence on black hole mass. Similarly, \citet{Selsing15} recently constructed a composite spectrum for very luminous quasars at $1 < z < 2.1$ from UV to near-infrared deriving a mean slope of $\alpha=-0.3$.

As anticipated in \figu\ref{fig|AlphaLoptMbh} the standard disk models (upper panels) with zero and maximal spin (magenta and green dotted lines) predict a steady reddening of \alf\ with increasing black hole mass at fixed luminosity.
The substantial uncertainties in virial black hole mass estimates of $\gtrsim 0.5$ dex \citep[e.g.,][and references therein]{ShenLiu12}, could however have a non-negligible impact on the comparison between models and data.
To probe this, we performed a series of Monte Carlo tests. We started by randomly extracting a very large number of black holes from the Type 1 corrected active black hole mass function at $z\sim 1$ by \citet{Schulze15}. To these we associated an Eddington ratio $\lambda=\log L_{\rm bol}/L_{\rm Edd}$ taken from a power-law with broad exponential cut-off distribution \citep[][]{Shankar13acc}, typical of AGN at these redshifts \citep[e.g.,][]{Bongiorno12}. We also experimented with different shapes for the underlying black hole mass function or Eddington ratio distribution without finding any strong change in the main outcomes of our tests. Bolometric luminosities were then converted to $L_{\rm 3000}$ adopting the quasar bolometric corrections by \citet{Runnoe12}, $\log L_{\rm bol}=0.75\times(0.97\log L_{\rm 3000}+1.85)$.
We then ran a few hundred realizations assigning at each iteration random Gaussian errors to black hole masses and luminosities with widths of $0.6$ dex and $0.1$ dex, respectively, and then computed the resulting mean \alf\ over all realizations as a function of black hole mass and luminosity.
The results of the test, marked by solid, magenta and green lines in the upper panels of \figu\ref{fig|AlphaMonteCarlo},
show that $\alpha_{3000}$ ``flattens out'', becoming significantly bluer at large black hole masses, aligning better with the data (though still predicting bluer slopes, especially at higher luminosities). This is expected given the significantly higher number densities of lower mass black holes coupled to the large errors in black hole masses assumed in the tests. These results agree well with those by \citet{Davis07} who also showed that large errors in black hole virial masses can effectively erase most of the dependence of UV slopes on black hole mass.

\begin{figure*}
    \center{\includegraphics[width=15truecm]{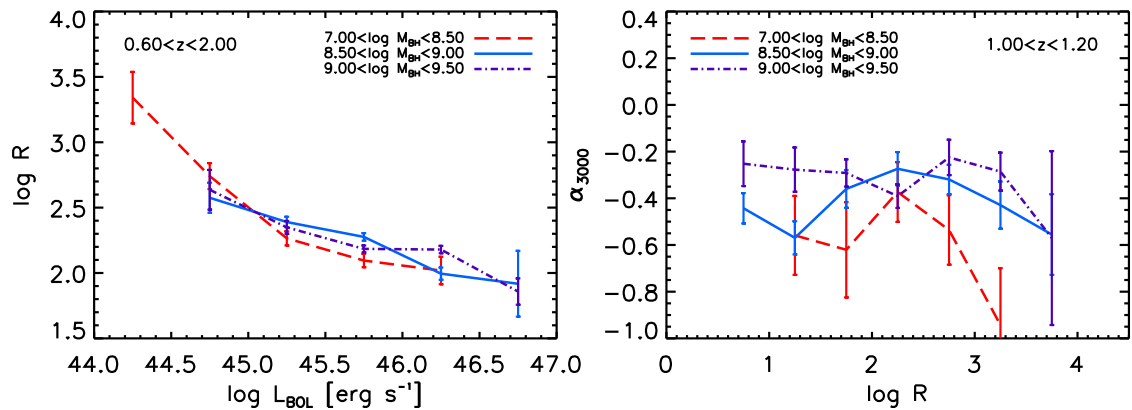}
    \caption{Left: Mean radio loudness as a function of bolometric luminosity (as reported by \citet{Shen11}) for different bins of black hole mass as labelled. Right: Mean \alf\ as a function of radio loudness $R$ for different bins of black hole mass, as labeled. While quasars become more radio-loud with decreasing luminosity, there is no clear dependence of \alf\ on $R$.
    \label{fig|AlphaRadioLoudness}}}
\end{figure*}

The lower panels of \figu\ref{fig|AlphaMonteCarlo} show the comparisons of our data with the AGNSPEC disk predictions with two extreme values of the input inclinations ($i=10^{\circ}$ and $i=70^{\circ}$, solid and long-dashed lines, respectively). We verified that models with
more realistic random input orientations fall within these two extreme cases. Models with low inclinations are in good agreement with those from basic blackbody models (solid lines in the upper panels), while the predicted \alf\ slopes tend to significantly decrease when assuming progressively higher inclinations. Interestingly, we identified a degeneracy among models characterized by low inclination/low spin and high inclination/high spin (purple solid and green dot-dashed lines, respectively). While high spin models show bluer \alf\ slopes due to the smaller inner disk radius, the redder UV slopes for viewing angles near the disk plane are mostly due to limb darkening effects.

We have also checked that including dust extinction in the predicted quasar spectra yields redder spectra further contributing to the degeneracy with inclination and spin. In particular, modeling the extinction curve as $A(\lambda)=0.06(5500\AA/\lambda)$ \citep[e.g.,][]{Capellupo15} yields $\Delta\alfe\sim0.1$.
Higher values of the extinction in quasars are also possible \citep[e.g.,][]{SternLaor12}. Rapidly spinning black holes need in general high inclinations $i \gtrsim 50^{\circ}$ with some degree of extinction to match the data. Black holes with retrograde spins may also yield redder slopes.

Other systematics may arise from adopting different input galaxy spectra. To check for this effect, we randomly selected quasar subsamples of different black hole mass and luminosity at $1<z<1.2$, and refitted the quasar spectra adopting the star-forming host galaxy template of Arp220 \citep[e.g.,][]{Polletta07}. Despite this template being more prominent in the UV with respect to our reference elliptical template (\sect\ref{sec|method}), we found resulting slopes on average bluer by a relatively modest $\Delta\alfe\sim 0.05-0.15$. This effect could anyway be significantly reduced by extinction.

We finally note that the complexities behind the exact modeling of absorption/emission features such as those from FeII/III groups, may limit the precision of our continuum measurements, though we do not expect them to significantly alter any of our results.

In \figu\ref{fig|AlphaRadioLoudness} we further investigate the connection between spectra and radio loudness in specifically radio-loud quasars.
There are clear trends (left panel) that the mean radio loudness $R$ decreases with increasing luminosity, and thus ultimately accretion rate \citep[e.g.,][]{Sikora07,Sikora13}. These trends hold irrespective of the exact black hole mass or redshift interval. However, the slope $\alpha_{3000}$ (right panel) shows no clear dependence on radio loudness $R$, at least for the radio-loud quasars in the interval $1<z<1.2$.

\section{Discussion}
\label{sec|discu}

Our analysis of quasar spectra at $z\sim 1$ pointed to rather constant and red mean values of the UV slope in the range $-0.5<\alfe<-0.3$ with a mild dependence on black hole mass. This (lack of) trend, we showed, is most probably induced by the significant uncertainties in black hole mass estimates (\figu\ref{fig|AlphaMonteCarlo}), while including high inclinations and extinction can align the predicted \alf\ to the redder slopes derived from the data.

We also find that the vast majority of radio-loud quasars have comparable or even redder \alf\ slopes with respect to optical quasars of similar black hole mass and accretion rate. This finding is of particular interest to the still unsolved issue of the origin of radio-loudness \citep[e.g.,][]{Sikora07,Kordi08,MerloniHeinz08,Shankar10radio,KratzerRichards}.
If radio-loud quasars are characterized by systematically higher spins than radio-quiet quasars, they would in fact be expected to show bluer UV slopes (\sect\ref{sec|intro}), in contrast to our data. Interestingly, \citet{Punsly14} has also reported a ``deficit'' in the UV spectra of radio-loud quasars
between 1100\ang\ and 580\ang, which might reflect what we observe in SDSS at longer wavelengths. If this spectral distortion is linked to local energy dissipation due to a large-scale magnetic flux \citep{Punsly14}, it should moderately scale with the spin and/or radio loudness, while no clear correlation is detected in our data between \alf\ and radio loudness (\figu\ref{fig|AlphaRadioLoudness}). Additional causes for a weak spectral dependence on spin and general spectral distortions may be related to AGN wind losses \citep[e.g.,][]{LD14,SloneNetzer}.

There are a number of cases in which the \citet{SSdisk} models have been successfully applied to describe quasar spectra
\citep[][]{Shields78,Kishimoto08,Capellupo15}. However, our results add to the mounting evidence for some disagreement between standard disk models and data derived from direct spectral fitting \citep[e.g.,][]{Bonning07,Davis07,SloneNetzer}, quasar microlensing \citep[e.g.,][]{Dai10}, or lags measured from correlated X-ray/UV/optical monitoring of AGN \citep[e.g.,][]{McHardy14}.

\section*{Acknowledgments}
We thank Ivan Hubeny and Eric Agol for their AGNSPEC and Kerrtrans9 codes and Omar Blaes for general help. We warmly thank Shane Davis and Ari Laor for very useful comments and for sharing their data. We also thank Gordon Richards for discussions and an anonymous referee for helpful comments that improved the clarity of the paper.


\label{lastpage}
\end{document}